\bigskip
\documentclass[preprint,12pt]{elsarticle}
\usepackage{graphicx}% Include figure files
\usepackage{fancyhdr}
\usepackage{dcolumn}% Align table columns on decimal point
\usepackage{amssymb}
\usepackage{amsmath}
\usepackage[normalem]{ulem}
\usepackage{bm}
\usepackage{textcomp}
  \usepackage{natbib}
  \usepackage[a4paper,left=1.55cm,top=2cm,right=1.55cm]{geometry}
\usepackage[colorlinks, linkcolor=red, anchorcolor=green,citecolor=blue ]{hyperref}
\usepackage[dvips]{color}

\journal{Physics Letters B}

\begin{document}

\begin{frontmatter}



\title{Strong correlation of the neutron star core-crust transition density with  the $\sigma$-meson mass via  vacuum polarization}


\author{Niu Li, Wei-Zhou Jiang, Jing Ye, Rong-Yao Yang, Si-Na Wei}

\address{School of Physics, Southeast University, Nanjing
211189, China}

\begin{abstract}
We study  the neutron star core-crust transition density $\rho_t$ with the inclusion of  the vacuum polarization in the dielectric function  in the nonlinear relativistic Hartree approach (RHAn).  It is found that the strong correlation between the $\rho_{t}$ and the scalar meson mass $m_{\sigma}$ strikingly overwhelms the uncertainty of the nuclear equation of state in the RHAn models, in  contrast to the usual awareness  that $\rho_{t}$ is predominantly sensitive to the isovector nuclear potential and symmetry energy. The  accurate extraction of $\rho_{t}$ through the future gravitational wave measurements  can thus provide a strong constraint on the longstanding  uncertainty of   $m_{\sigma}$, which is of significance to better infer the vacuum property. As an astrophysical implication, it suggests that  the correlation between $\rho_t$ and $m_\sigma$ is very favorable to reconcile the difficulty in reproducing the large  crustal moment of inertia for the pulsar glitches with the well constrained symmetry energy.
\end{abstract}


\end{frontmatter}

\section{Introduction}

A transition from the homogeneous core matter to inhomogeneous crust in neutron stars can result
in a series of phenomena. The neutron star radius \cite{ref1}, pulsar glitches \cite{ref2,ref3,
ref4,ref5,ref6}, crust meltdown in inspiraling binary neutron stars \cite{ref7}, asteroseismology from
giant magnetar flares \cite{ref8}, crust relaxation in cooling and accreting neutron stars \cite{ref9},
and so on are all closely related to the core-crust transition density. The determination of the
core-crust transition density is thus an important issue of broad interest. Without precision measurement available at present,  obtaining the $\rho_t$  is rather dependent on    various approaches such as the thermodynamical~\cite{ref10,ref11,
ref12,ref13,ref51,ref52,ref53} and hydrodynamic~\cite{Ch04,Du07,Xu09,Zh12}  methods and the relativistic random phase approximation
(RRPA)~\cite{ref23,ref24,ref26}. Regardless of the uncertainty in the obtained $\rho_t$, the common problem  of these approaches is the negligence of  the vacuum contribution.  In reality, the importance of the vacuum  not only exists in the spontaneous breaking of chiral symmetry for the mass acquisition of the hadrons~\cite{ref31,Sc01,Mao10,ref32,ref33,ref34} but is manifested in the bulk properties such as  the nucleon effective mass~\cite{ch77, Se86}, compression modulus~\cite{glen88,ji88,wa21},  chiral condensate in nuclear matter~\cite{jia08} and the giant resonances in finite nuclei~\cite{ref27,ref28,ref29}. Since the vacuum is tightly associated with the matter through various vacuum polarizations and  condensates (e.g., the $\sigma$ mode in the chiral model), it is inevitably of importance  to ask and answer whether  the vacuum contribution is hidden in the deep background for the low-density phenomenology or  deserves a rigorous check.

In nuclear many-body systems characteristic of the Fermi sea and surface with the interaction mediated by the mesons on the hadronic level, the vacuum effect arises predominantly from  the scalar mode or $\sigma$ meson, as is manifested in the relativistic Hartree approximation (RHA)~\cite{ch77},  the linear $\sigma$ model~\cite{ref31} and its linear~\cite{Sc01,Mao10} and nonlinear~\cite{We68,We79,Ko97}  extensions where the vacuum all situates on the circle $<\sigma>=f_\pi$,  the
dilaton model of the spontaneous breaking of the scale invariance~\cite{ref37}, and  the Nambu-Jona-Lasinio  model \cite{ref32,ref33} as well.  Consistent with these various sources, the existence of $\sigma$ meson  has been controversial, since the $\sigma$ meson was proposed
more than 60 years ago~\cite{ref38,ref39}. The mass measurements also  exhibit a large  width remaining roughly from  400 to 650 MeV nowadays~\cite{ref38,Wo22},  which can typically be  treated as a composite $\pi\pi$ resonance~\cite{ref35,ref36,Bri17}.
As the $\sigma$ meson usually defines the  spontaneous breakdown of the vacuum, the identification of the $\sigma$ meson and its mass $m_\sigma$ actually is of the fundamental importance. In deed, the $\sigma$ meson mass  is highly relevant to the vacuum through the renormalization of the scalar polarization. In nuclear medium,  the in-medium polarizations may be expected to encode the constraint on the  $m_\sigma$. It is thus the aim of this Letter to dig out the relationship between $m_\sigma$ and  $\rho_t$ by exploring  the medium-vacuum dual property  of the in-medium polarizations.
We will show that the scalar vacuum polarization induces a strong correlation  between $m_\sigma$ and  $\rho_t$ that is very beneficial in determining the $m_\sigma$ and  $\rho_t$ both by the experiments designed for measuring one of them~\cite{ref7,Wo22,Abb17}. In addition, the implication of the strong correlation in the glitches of pulsars~\cite{An12,Ch13} is addressed.

\section{Formalism}

The lagrangian density, based on the Walecka model~\cite{Se86},  includes the free fields  of nucleons and mesons,  the interaction between them, and the nonlinear $\sigma$ self-interactions~\cite{Bo77}.  The interacting lagrangian for the latter two parts reads
\begin{eqnarray}
\begin{aligned}
\mathcal{L}_{\rm int}=&\overline{\psi} [\gamma_{\mu}(i\partial^{\mu}-g_{\omega}V^{\mu} -g_{\rho}
             \boldsymbol{\tau\cdot b^{\mu}})
             -(M-g_{\sigma}\phi)]\psi-U(\phi),
\end{aligned}
\end{eqnarray}
where  $U(\phi)=g_{2}\phi^{3}/3!+g_{3}\phi^{4}/4!$ is    the nonlinear $\sigma$ meson self-interactions. In this work, we follow
the relativistic Hartree approximation (RHA) to treat the mean fields and polarizations self-consistently~\cite{ch77}.
The divergent terms in the nucleon self-energy and meson polarizations are renormalized by introducing the consistent counterterms $\mathcal{L}_{CT}=\sum^{4}_{n=1}\frac{\alpha_{n}}{n!}\phi^{n}$~\cite{ch77}. In addition, the one-loop contributions from the $g_2$ and $g_3$ terms need to be renormalized by introducing the additional counterterms. Eventually,  in RHA the renormalized finite $U^{Ren.}(\phi)$ is given as~\cite{Se86}
\begin{eqnarray}
\begin{aligned}\label{eqUren}
U^{Ren.}(\phi)=&\frac{1}{3!}g_{2}\phi^{3}+\frac{1}{4!}g_{3}\phi^{4}+ \frac{1}{(8\pi)^{2}}
    [(m^{2}_{\sigma}+g_{2}\phi+\frac{g_{3}\phi^{2}}{2})^{2}\ln(1+\frac{g_{2}\phi}{m^{2}_{\sigma}}
       +\frac{g_{3}\phi^{2}}{2m^{2}_{\sigma}})\\
    &- m^{2}_{\sigma}(g_{2}\phi+\frac{g_{3}\phi^{2}}{2})-\frac{3}{2}(g_{2}\phi+\frac{g_{3}\phi^{2}}{2})^{2}
    -\frac{(g_{2}\phi)^{2}}{3m^{2}_{\sigma}}(g_{2}\phi
    +\frac{3g_{3}\phi^{2}}{2})+\frac{(g_{2}\phi)^{4}}{12m^{4}_{\sigma}}].
\end{aligned}
\end{eqnarray}
Here, the introduction of the nonlinear terms in $U(\phi)$ is necessary for reducing the compression modulus to the empirical region, and the additional terms in Eq.(\ref{eqUren}) arising from the renormalization of $U(\phi)$ affect the specific parametrizations constrained by  the saturation properties.

In the propagation of the  nucleon-nucleon interaction, the whole polarization tensor in the dressed meson propagator, the density-dependent and Feynman parts, should be
treated on the equal footing. This is  the principle  in  treating the dielectric function whose zero points are the centroid of the collective modes. In a static potential with $q_0=0$, the sign of the dielectric function $\epsilon_L$ is used to judge the stability of uniform neutron star matter below saturation density. The transition density $\rho_t$ from stable to unstable matter is the largest density determined by the relation
\begin{eqnarray}
\begin{aligned}\label{eq3}
\epsilon_L=\det[1-D_{L}(q)\Pi_{L}(q)]_{q_0=0}\leq 0,
\end{aligned}
\end{eqnarray}
for any $q$~\cite{ref23,ref24,ref26}. In uniform matter with protons, neutrons, electrons and muons at beta equilibrium, the meson propagator $D_L$ and polarization tensor $\Pi_L$ are   $4\times4$ matrices~\cite{ref26}, and all the single entries  in $\Pi_L$  include the finite Feynman parts, among which  the Feynman part of the scalar  polarization contributes dominantly to shifting the transition density.

\section{Results and discussion}

\begin{figure}[!htb]
\centering
\includegraphics[height=7.5cm,width=8cm]{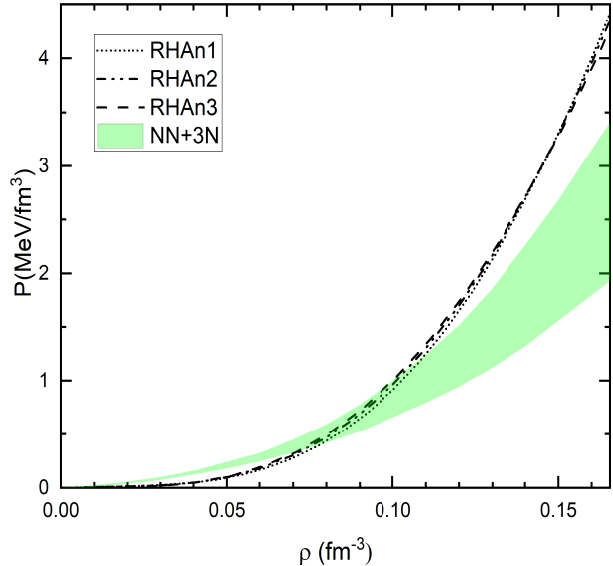}
\caption{The pressure of pure neutron matter with RHAn parametrizations.  The shaded area represents  the result with the chiral NN+3N interactions.}
\label{figc}
\end{figure}

The early RHA model is the renormalized version of the Welacka model~\cite{ch77}, whose compression modulus of 461.1 MeV is still very high.  With the inclusion of the  nonlinear $\sigma$ meson self-interactions in Eq.(\ref{eqUren}) to reduce the compression modulus, we obtain the specific RHA parametrizations constrained by the saturation properties  of nuclear matter: $\rho_{0}=0.16$ $fm^{-3}$,  the binding energy per nucleon
$E_{b}=-16$ MeV, the vanishing pressure at $\rho_0$, and the given incompressibility. The coupling constant $g_\rho$ of the $\rho$ meson is adjusted to meet some average of a variety of constrained values of the symmetry energy at saturation density. All the meson masses but  $m_\sigma$ are given at their experimental values.
In Table~\ref{tab1}, we tabulate a few selected   RHA parametrizations and the corresponding predictions of nuclear matter at given $m_\sigma$ (512 MeV). It is seen from Table~\ref{tab1} that the compression modulus becomes reasonably soft in the nonlinear RHA (RHAn) parametrizations together with the increase of the nucleon effective mass~\cite{glen88}. Note that the current data analysis can constrain the incompressibility with an upper bound of  315 MeV~\cite{St14,Ro18}. In addition, we  mention that  the  pressure of pure neutron matter obtained with these RHAn parametrizations is basically consistent with that calculated with the microscopic chiral NN and 3N interactions~\cite{He13} at low density and some deviation arises at the  density above 0.11$fm^{-3}$, as shown in Fig.~\ref{figc}.

\begin{table}[h]
\centering
\begin{tabular}{l|c|c|c|c|c|c|c|c|c}
  \hline \hline
 Model  &$g_{\sigma}$ & $g_{\rho}$ &$g_{\omega}$& $g_{2}$ & $g_{3}$&$M^*/M$&$\kappa$ &$\rho_{t}$  &    $P_{t}$     \\ \hline
 RHA  &7.99   & 3.91       &9.79         & 0              &  0     & 0.726&461.1  &0.078      & 0.265           \\ \hline
 RHAn1 &7.70   & 4.03      &8.39         & 27.9           &  -29.5  &    0.784&300.0  &0.081      & 0.450           \\ \hline
  RHAn2&7.51   & 4.09       &7.71         & 40.8           &  -45.2&    0.809&270.0  &0.080      & 0.477             \\   \hline
  RHAn3&7.11   & 4.15       &6.51         & 61.1           &  -75.7&    0.847&240.0  &0.078      & 0.459            \\
\hline
\end{tabular}
\caption{Various RHA parametrizations and some properties of nuclear matter without (i.e., RHA) and with the inclusion of the  $\sigma$-meson self-interacting terms.  $\kappa$, $\rho_t$ and $P_t$ are the incompressibility (MeV), transition density ($fm^{-3}$) and pressure (MeV$/fm^3$), respectively. Here, the masses of $\sigma$, $\omega$, and $\rho$ mesons are 512, 783, and 770 MeV, respectively.\label{tab1} }
\label{tab:Margin_settings}
\end{table}

\begin{figure}[thb]
\centering
\includegraphics[scale=0.60]{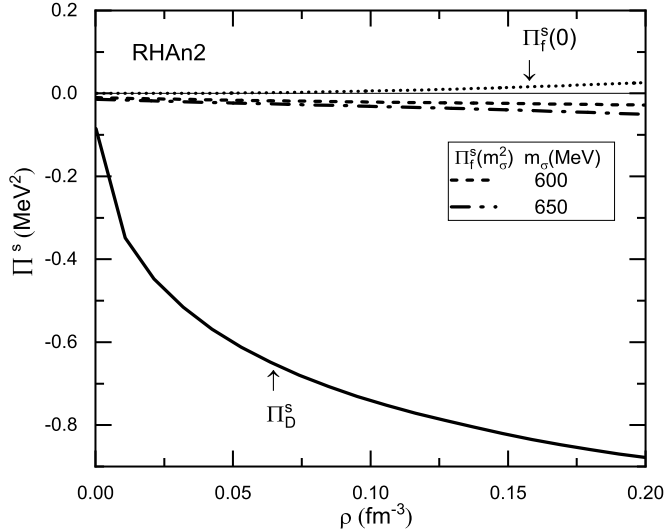}
\caption{The $\Pi^{s}_{f}(q^2)$ and  $\Pi^{s}_{D}$ as a function of density.  Three curves of $\Pi^{s}_{f}(q^2)$ are evaluated at  $q^2=m^2_\sigma$ ($m_\sigma=600,650$ MeV) and 0, respectively. A thin solid horizontal line is drawn for reference. }
\label{fig1}
\end{figure}

\begin{figure}
\centering
\includegraphics[scale=0.60]{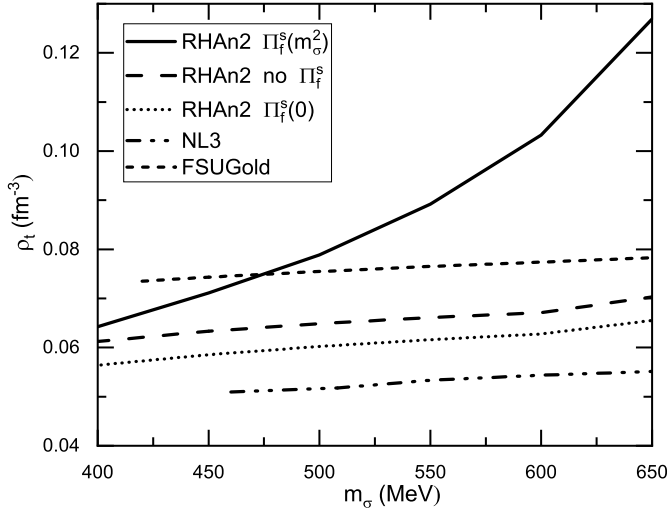}
\caption{The transition density $\rho_{t}$ versus $m_\sigma$.   Various curves stand for  three cases of  $\Pi^{s}_{f}(q^2)$ in RHAn2  and  two RMF results without $\Pi^{s}_{f}$, respectively.}
\label{fig2}
\end{figure}

\begin{figure}[thb]
\centering
\includegraphics[scale=0.60]{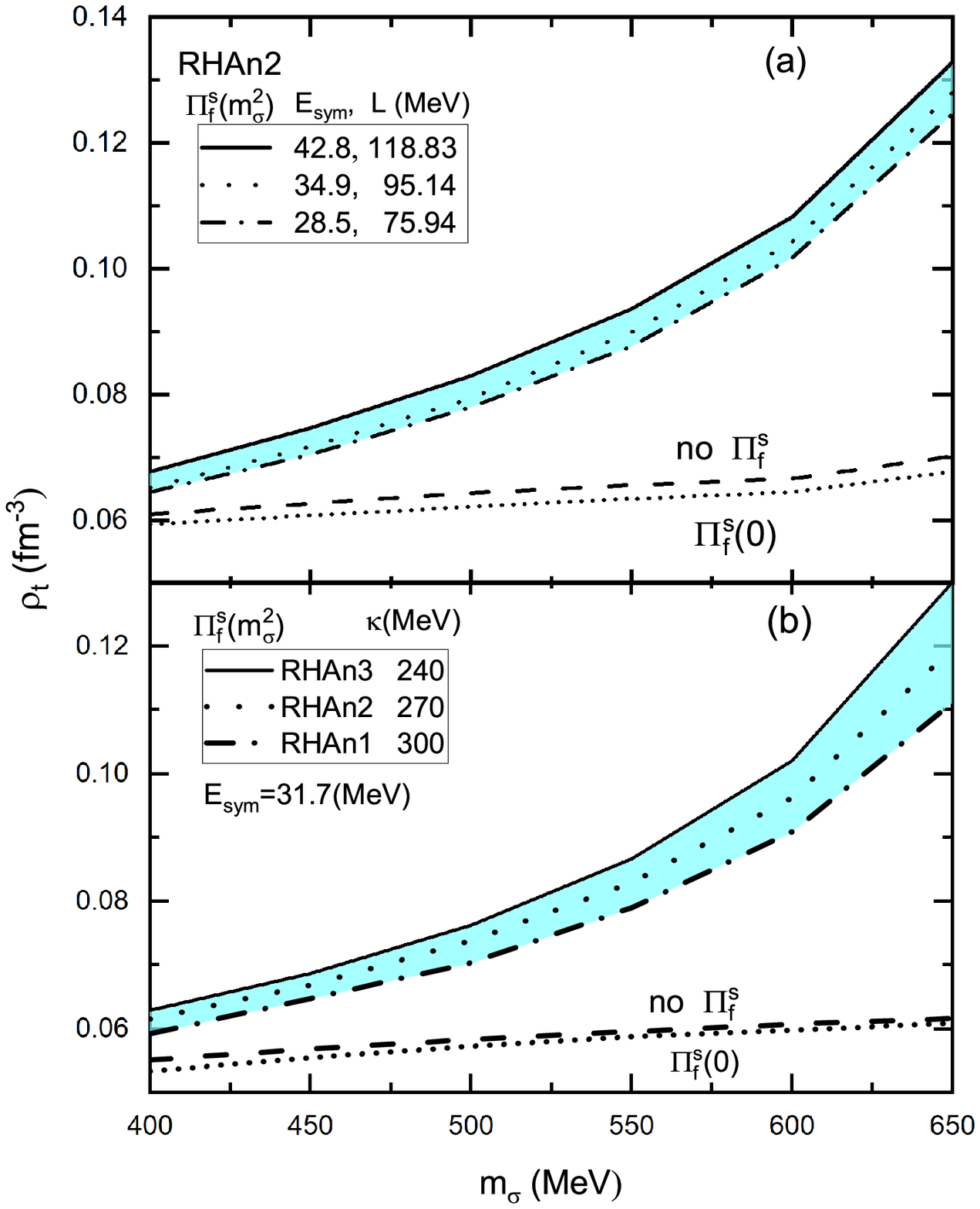}
\caption{(color online) $\rho_{t}$ versus $m_\sigma$ with the effect of the EOS uncertainty: The influence of the symmetry energy (a) and of the incompressibility (b). In (a), $L$ is the slope parameter of the symmetry energy. }
\label{fig3}
\end{figure}

%The effect of the RHAn model pressure growth rate higher than the chEFT result is that the maximum neutron star mass is in equilibrium with the attraction when it is relatively small, i.e. only the maximum mass of the RHAn1 model is barely more than twice the solar mass.

In the renormalization of $\Pi^{s}_{f}(q^2)$, we  choose the natural renormalization point on the mass shell $q^{2}=m_\sigma^2$  as in Ref.~\cite{ref42}.
Different from the renormalization point $q^{2}=0$ in Refs.~\cite{ch77,Rj88},
the on-shell point $q^{2}=m^{2}_{\sigma}$ actually redefines the physical mass of the $\sigma$ meson in the propagator renormalization. Fig.~\ref{fig1} shows that the change trend of $\Pi^{s}_{f}(m^2_\sigma)$ is opposite to  $\Pi^{s}_{f}(0)$, and $\Pi^{s}_{f}$ in both cases are small compared
with $\Pi^{s}_{D}$.

Exhibited in Fig.~\ref{fig2} is the transition density $\rho_{t}$  as a function of  $m_{\sigma}$. It is seen that the  $\rho_{t}$ with $\Pi^s_f(q^2)$ renormalized at
$q^{2}=m^{2}_{\sigma}$  increases strikingly with the rise  of
$m_{\sigma}$. In comparison,
$\rho_{t}$ in the cases with $\Pi^s_f(q^2=0)$   and without $\Pi^s_f$ varies very slightly with the rise of $m_{\sigma}$, since there is no $m_\sigma^2$ dependence in  $\Pi^s(q^2)$.
We also calculate $\rho_t$ with the usual RMF models NL3 and FSUGold,  and the variation of $\rho_t$ versus $m_\sigma$ is similarly even.  In these cases, the small dependence on $m_\sigma$ arises from the $\sigma$ meson propagator and the coupling constant $g_\sigma$ that is readjusted at various $m_\sigma$ to keep the saturation properties unchanged. Given the saturation properties, the relationship between $\rho_{t}$ and $m_\sigma$ is unique.
Though the difference between $\Pi^{s}_{f}(q^2)$ at $q^2=0$ and $m_\sigma^2$, shown in Fig.~\ref{fig1},
is  small,  their effects on the $\rho_{t}$ depart to be in sharp contrast.  The underlying reason lies in the subtle cancelation in obtaining the zero points of the dielectric function $\epsilon_L$ in Eq.(\ref{eq3}). This can be observed in a simplified $\epsilon_L$ by neglecting the Coulomb interaction and photon polarization
\begin{eqnarray}
\begin{aligned}\label{eq4}
\epsilon_L=&[1+d_{s}(\Pi^{s}_{D}+\Pi^{s}_{f})]\cdot
[1+4d_{\omega}d_{\rho}\Pi^{2}_{L}+2(d_{\omega}+d_{\rho})\Pi_{L}]
-4(1+2d_{\rho}\Pi_{L})d_{s}d_{\omega}\Pi^{2}_{M},
\end{aligned}
\end{eqnarray}
where $ d_{s}=g^{2}_{\sigma}/(m^{*2}_{\sigma}-q^{2})$, $d_{\omega}=g^{2}_{\omega}/(m^{2}_{\omega}-q^{2})$, and $
d_{\rho}=g^{2}_{\rho}/(m^{2}_{\rho}-q^{2})$, with $m_\sigma^*$ being the $\sigma$ meson effective mass~\cite{ref26}. The zeros are the result of the cancelation of the two terms in Eq.(\ref{eq4}). With the increase of $m_\sigma$, the second term diminishes due to the smaller $d_s$. The different sign of $\Pi_D^s$ and $\Pi_f^s$ in the first term is favorable to keep the equilibration in the cancelation of the two terms, which is the case of the $\Pi^s_f(q^2=0)$, see Fig.~\ref{fig1}. Thus,
the negative value of $\Pi^{s}_{f}$, which is of the same sign of   $\Pi^{s}_{D}$, shifts the zero point of the dielectric function to a  larger density, that is, the transition density $\rho_{t}$ goes up clearly with the rise of $m_\sigma$. We stress that the sensitive dependence of $\rho_t$ on $m_\sigma$ arises from the appropriate inclusion of the Dirac sea contribution and is therefore absent in the usual  RMF models, see Fig.~\ref{fig2}. Note that the range of $m_\sigma$ in Figs.~\ref{fig2} and \ref{fig3} is taken as 400-650 MeV, according to data from the particle data group PDG2022~\cite{Wo22}.

%\begin{figure}
%\centering
%\includegraphics[scale=0.35]{ImomVSMsigma.png}
%\caption{(color online) Fraction of crustal moment of inertia as a function of neutron star mass.
%   Fig(a) shows the minimum $m_{\sigma}$ required to reach  the necessary
%fraction at $E_{sym}$=33MeV and a symmetric energy slope L= 89.43MeV; In Fig(b) we use $E_{sym}$=42.8MeV and L= 118.83MeV. }
%\label{fig4}
%\end{figure}

As the dielectric function and transition density is tightly associated with the EOS of asymmetric matter through the meson propagators (nuclear potentials), it is necessary to estimate the blurring effect, arising from the EOS uncertainty mainly in the symmetry energy and compression modulus~\cite{Ro18,Li08,Jm16},   on the $\rho_t-m_\sigma$ relationship. We depict the distribution domains of $\rho_{t}$ corresponding to some bounds of the symmetry energy and incompressibility  in
Fig.~\ref{fig3}. The symmetry energy at saturation density seems to be  well constrained with a small uncertainty of $31.7\pm3.2$ MeV~\cite{Ro18,ref43a}, while its scatter can roughly range from 28.5 to 42.8 MeV in various models~\cite{ref43,ref44,ref45,ref47,ref48,ref49,ref50}. We see from Fig.~\ref{fig3}a (the upper panel) that a considerable variation of the symmetry energy from 28.5 to 42.8 MeV in RHAn2 (with $\kappa=270$ MeV) just leads to a moderate shift of $\rho_t$. Here, the effect of the slope parameter of the symmetry energy ($L=3\rho_0\partial E_{sym}/\partial \rho$ at saturation density $\rho_0$) on $\rho_t$ is implicit, as the slope is positively relevant to the symmetry energy through the same parameter $g_\rho$. In this case, the apparent relationship between $\rho_t$ and $L$ is opposite to that in the literature~\cite{ref51,ref52,ref53}, while the consistency can be reproduced by invoking the additional term of the isoscalar-isovector coupling to modify the $\rho$ meson effective mass as in Ref.~\cite{ref24}. 
The slope parameter in our models ranges from about 70 to 120 MeV, which is roughly consistent with $106\pm37$ MeV, extracted from the PREX-2 experiment~\cite{PREX}. In fact, the slope parameter extracted from  various experiments is not very consistent, for instance, see
Refs.~\cite{ref47,ref50,PREX,Roca15,ref54,ref55}. In Refs.~\cite{ref51,ref52,ref53},  a large slope range of 20-120 MeV is used to delimit the variation of $\rho_t$ around 0.08-0.12 $fm^{-3}$ that  clearly exceeds the shaded region shown in Fig.~\ref{fig3}a. This suggests that the more accurate experiments to extract the $L$ and $m_\sigma$ are of similar importance.

In Fig.~\ref{fig3}b,  given  the  symmetry energy at 31.7 MeV, we change the incompressibility reasonably from 240 to 300 MeV. It can be seen  that the transition density is also moderately affected by the compression modulus.  Consequently, the moderate effect of the EOS in the present nonlinear RHA approach reconfirms the fact that the variation of $\rho_{t}$ is decided overwhelmingly  by  $m_{\sigma}$ through the vacuum polarization. Concerning the disparity  of the $\rho_t$ in various models or approaches~\cite{ref6,Zh12,ref24,ref26},  the experimental extraction of $\rho_t$ not only calibrates the various models, but can  further be used to constrain the large uncertainty of $m_\sigma$ for its high sensitivity to $\rho_t$ together with the well constrained EOS. As  the third-generation gravitational wave detector network is in the plan with high precision~\cite{Abb17}, it is  very
likely  to detect the gravitational wave phase transition induced by  inspiraling binary neutron stars and thus obtain the transition density $\rho_{t}$~\cite{ref7}. Once $\rho_{t}$ is extracted from the gravitational  wave data in the future, $m_{\sigma}$ can be well constrained. Besides, the constraint on the $\rho_t$ extracted indirectly from various terrestrial experiments is similarly of significance~\cite{Mo10}.  Here, we should mention that the restriction of $m_\sigma$  with the bulk matter models for finite nuclei is not straightforward, since the value of $m_\sigma$ is dependent on how to include the in-medium effects and high-order correlations in various approximations. Even in the mean-field approximation,  the value of $m_\sigma$ cannot be determined solely but dependent on  the $\sigma$ meson self-interactions in terms of $g_2$ and $g_3$, see Tables~\ref{tab1} and  \ref{tab2}.

\begin{table}[h]
\centering
\begin{tabular}{l|c|c|c|c|c|c|c|c|c|c}
  \hline \hline
  Model&$m_{\sigma}$&$g_{2}$ & $g_{3}$   & $\rho_{t}$ & $L$  & $E_{sym}$&$M_{max}$&$R_{1.4}$& $\Lambda_{g}$  & $\Delta I/I$         \\ \hline
  RHAn1&550.0       &  33.9  &  -28.5    & 0.088     & 90.37 &  33.0  &  2.04    &13.29&  667.6      &    7.08               \\ \hline
  RHAn1&454.0       &  20.3  &  -29.4    & 0.075     & 119.80 &  42.8  &  2.03   &14.51&  711.1      &    7.02               \\ \hline
  RHAn2&555.0       &  50.5  & -43.8     & 0.090     &  89.45 &  33.0  &  1.92   &13.00&  560.5      &    7.09               \\ \hline
  RHAn2&471.0       &  32.7  & -44.5     & 0.078     & 118.87 &  42.8  &  1.91   &14.28&  597.2      &    7.00               \\ \hline
  RHAn3&580.0       &  83.7  & -70.1     & 0.095     &  88.13 &  33.0  &  1.73   &12.48& 408.5      &    7.06               \\ \hline
  RHAn3&514.0       &  61.7  & -75.7    & 0.084      & 117.55 &  42.8  &  1.72   &13.88& 438.0      &    7.01               \\
\hline
\end{tabular}
\caption{ Quantities constrained by the $\Delta I/I$ (around 7\% for $1.4M_\odot$ pulsars) and $E_{sym}$ for various models.  $m_{\sigma}$, $E_{sym}$ and its slope $L$ are in units of MeV, $\Lambda_g$ is the dimensionless tidal deformability of the 1.4 $M_\odot$ neutron star with its radius $R_{1.4}$, and the maximum neutron star mass is in units of $M_\odot$. \label{tab2} }
\label{tab:Margin_settings}
\end{table}

To associate astrophysical data with $\rho_t$ and $m_\sigma$ based on the $\rho_t-m_\sigma$ correlation,  here we evaluate the fraction of crustal inertia moment for the neutron star glitches. For neutron stars, the Tolman-Oppenheimer-Volkoff  equation is solved with the  EOS of asymmetric matter as an input. To reproduce the glitches of the Vela pulsar,  the proportion of crustal moment of inertia $\Delta I/I$  should increase  significantly from $1.6\%$ to $7\%$ due to  the crustal entrainment that impairs the neutron superfluid~\cite{An12,Ch13}.
 Given the mass of a neutron star, the   $\Delta I/I$  is decided by the transition density $\rho_{t}$ and star radius both of which have a clear dependence on the symmetry energy, and its large value of  $7\%$ requires usually a large $\rho_{t}$ and a stiff symmetry energy.
In Table~\ref{tab2}, we tabulate the lower limit of $\rho_t$ and the corresponding $m_\sigma$, given the symmetry energy of $E_{sym}=33$ and 42.8 MeV and the value of $\Delta I/I$ around 7\% for the 1.4$M_\odot$ neutron star.
According to the $\rho_{t}-m_{\sigma}$ correlation, the lower limit of $\rho_t$ corresponds to a minimum $m_\sigma$ in Table~\ref{tab2} required to reach  $\Delta I/I$  of 7\% under different symmetry energies.
It can be seen from Table~\ref{tab2} that for the symmetry energy from 33 to 42.8 MeV with its slope $L$ roughly from 90 to 118 MeV,
 the $\rho_{t}$ is necessarily reduced to reach  $\Delta I/I$  of 7\% due to the increase of the NS radius~\cite{ref6}.
On the other hand, once the symmetry energy that is independent of $m_\sigma$ is well constrained, the appropriate inclusion of the vacuum polarization actually opens a new window to meet the required fraction  $\Delta I/I$  simply through $\rho_t$ by adjusting $m_\sigma$. This is also of practical sense especially when it is somehow difficult for some models to fit the data of the Vela pulsar~\cite{ref6}.

In Table~\ref{tab2}, we also tabulate the tidal deformability $\Lambda_g$~\cite{Sh21} whose experimental bounds for the  1.4$M_\odot$ neutron star are $190^{+390}_{-120}$, extracted from the gravitational wave data~\cite{Ab18}. Here, $\Lambda_g$ is mostly affected by the symmetry energy which dominates the uncertainty of the neutron star radius, while the transition density $\rho_t$ just has a moderate effect on $\Lambda_g$. The maximum neutron star mass in Table~\ref{tab2} is positively (negatively) correlated with the incompressibility (nucleon effective mass), see Table~\ref{tab1}.
It is pointed out in Ref.~\cite{We11} that the maximum mass of the neutron star increases with decreasing  $m_\sigma$  for given $g_{\sigma}$.  Our models can present similar results, whereas the variation domain of $m_{\sigma}$ for a larger $M_{max}$ is rather limited in the nonlinear RHA model, since the nucleon effective mass can not be reduced significantly in a limited variable domain of the logarithmic function.
The maximum masses with the RHAn2 and RHAn3,  less than the experimental bound 2$M_\odot$~\cite{De10,An13}, can be improved by moderately reducing the nucleon binding energy $E_b$. With $E_b$ at the lower bound  ($-15$ MeV), the $M_{max}$ with the RHAn2 and RHAn3 are increased to be $1.98M_\odot$ and $1.84M_\odot$, respectively. Here, in addition to the incompressibility, a key factor to increase $M_{max}$ is the conversion  of the nucleon effective mass $M^*$ into the energy and pressure to be against gravitational collapse by dropping $M^*$. Since the nonlinear RHA models are not able to have a much smaller  $M^*$, we introduce the new term $g_{\sigma\omega}\sigma^2\omega_\mu\omega^\mu/2$~\cite{Hai08,Sha09} in the renormalizable formalism to manage shifting the $M^*$ down. A moderate rise of $M_{max}$ with some reduction of $M^*$ can be obtained to be about 0.01, 0.04, and 0.11 $M_\odot$ for RHAn1, RHAn2 and RHAn3, respectively.

\section{Summary}
In this work, we study the relationship between the $\sigma$-meson mass $m_{\sigma}$ and the core-crust transition density $\rho_{t}$ of neutron stars with the inclusion of the renormalized  vacuum polarization in the RRPA. In the presence of the large variations  of $m_\sigma$ and $\rho_t$ both, we have obtained a strong correlation between $m_\sigma$ and $\rho_t$  overwhelming  the uncertainty of the nuclear EOS in the RHAn models, which adds new content to the usual awareness that the $\rho_{t}$ is dominated by the isovector nuclear potential and symmetry energy.   The surprisingly strong dependence of $\rho_t$ on the $m_{\sigma}$  results from  the vacuum contribution of the  scalar polarization that modulates sensitively the zeros of the dielectric function. This strong correlation provides a vigilant reminder that the ubiquitous vacuum can not be easily ignored and has the experimental sense.
Once the  $\rho_{t}$ is determined  through the future advanced gravitational wave detectors, one can use it to  delimit the value of $m_{\sigma}$, and vice versa if the  $m_{\sigma}$ can be measured accurately by the terrestrial experiments. An astrophysical application of the correlation is that it is easy to  achieve the large  fraction (7\%) of the crustal moment of inertia necessary for the Vela pulsar glitches by moderately increasing the value of $m_\sigma$, without altering  the existing constraints on the EOS.

\section*{Acknowledgements}
This work was supported in part by the National Natural Science Foundation of China under Grant No. 11775049.


\begin{thebibliography}{00}


\bibitem{ref1}M. Fortin, C. Provid\^encia, A. R. Raduta, F. Gulminelli, J. L. Zdunik,
              P. Haensel, and M. Bejger, Phys. Rev. C \textbf{94}  (2016) 035804.
\bibitem{ref2}J. M. Lattimer and M. Prakash, Phys. Rep. \textbf{442}  (2007) 109.
\bibitem{ref3}N. Chamel, Phys. Rev. C \textbf{85}  (2012) 035801.
\bibitem{ref4}N. Andersson, K. Glampedakis, W. C. G. Ho, and C. M. Espinoza, Phys. Rev.
              Lett. \textbf{109}  (2012) 241103.
\bibitem{ref5}J. Piekarewicz, F. J. Fattoyev, and C. J. Horowitz, Phys. Rev. C \textbf{90}  (2014) 015803.
\bibitem{ref6}S-N. Wei, R-Y. Yang, W-Z. Jiang, Chin. Phys. C \textbf{42}  (2018) 74.
\bibitem{ref7}Z. Pan, Z. Lyu, B. Bonga, N. Ortiz, and H. Yang, Phys. Rev. Lett. \textbf{125}  (2020) 201102.
\bibitem{ref8}A. T. Deibel,  A. W. Steiner, and E. F. Brown, Phys. Rev. C \textbf{90}  (2014) 025802.
\bibitem{ref9}S. Lalit, Z. Meisel, and E. F. Brown. Astrophys. J. \textbf{882}  (2019) 2.
\bibitem{ref10}F. Douchin and P. Haensel, Phys. Lett. B \textbf{485}  (2000) 107.

\bibitem{ref11}B. A. Li, A. T. Sustich, M. Tilley and B. Zhang, Nucl. Phys. A \textbf{699}  (2002) 493.


\bibitem{ref12}S. Kubis, Phys. Rev. C. \textbf{76}  (2007) 025801.

\bibitem{ref13}Ch. C. Moustakidis, T. Niksi\'c, G. A. Lalazissis, D. Vretenar, and
              P. Ring, Phys. Rev. C \textbf{81}  (2010) 065803.

\bibitem{ref51}C. Ducoin, J. Margueron, and  C, Provid\^encia,  Europhys. Lett. \textbf{91}  (2010) 32001.
\bibitem{ref52}W. G. Newton, M. Gearheart, and B. A. Li,  Astrophys. J. Suppl. Ser. \textbf{204}  (2013) 9.
\bibitem{ref53}C. Ducoin, J. Margueron, C. Provid\^encia, and I. Vida\~na,  Phys. Rev. C \textbf{83}  (2011) 045810.


\bibitem{Ch04}P. Chomaz, M. Colonna, and J. Randrup, Phys. Rep. \textbf{389}  (2004) 263.
\bibitem{Du07}C. Ducoin, Ph. Chomaz, and F. Gulminelli, Nucl. Phys. A \textbf{789}  (2007) 403.
\bibitem{Xu09}J. Xu, L-W. Chen, B. A. Li and H-R. Ma, Astrophys. J. \textbf{697}  (2009) 1549.

\bibitem{Zh12}H. Zheng and L. W. Chen,  Phys. Rev. D  \textbf{85}  (2012) 043013.




\bibitem{ref23}K. Lim and C. J. Horowitz, Nucl. Phys. A \textbf{501}  (1989) 729.
\bibitem{ref24}C. J. Horowitz and J. Piekarewicz,  Phys. Rev. Lett.  \textbf{86}  (2001) 5647.

\bibitem{ref26}J. Carriere, C.J. Horowitz, and J. Piekarewicz, Astrophys. J. \textbf{593}   (2003) 463.

\bibitem{ref31}M. Gell-Mann, M. Levy, Nuovo Cim. \textbf{16}  (1960) 705.
\bibitem{Sc01}    O. Scavenius, \'A. M\'ocsy, I. N. Mishustin, and D. H. Rischke, Phys. Rev. C \textbf{64}  (2001) 045202.
\bibitem{Mao10}H. Mao, J. Jin, and M. Huang, J. Phys. G \textbf{37}   (2010) 035001.
\bibitem{ref32}Y. Nambu, G. Jona-Lasinio, Phys. Rev. \textbf{122}  (1961) 345.

\bibitem{ref33}T. Hatsuda, T. Kunihiro, Phys. Rep. \textbf{247}  (1994) 221.
\bibitem{ref34}S. N. Wei, W. Z. Jiang, Y. R. Yang, and D. R. Zhang, Phys. Lett. B \textbf{763}  (2016) 145.


\bibitem{ch77}S. A. Chin, Ann. Phys.  \textbf{108}  (1977) 301.

\bibitem{Se86}D. B. Serot and J. D. Walecka, Adv. Nucl. Phys. \textbf{16}  (1986) 1.

\bibitem{glen88}N. K. Glendenning, Phys. Lett. B \textbf{208}  (1988) 335.

\bibitem{ji88}X. Ji, Phys. Lett. B \textbf{208}  (1988) 19.
\bibitem{wa21}S. Wang, Q. Zhao, P. Ring, J. Meng,  Phys. Rev. C \textbf{103}  (2021) 054319.

\bibitem{jia08}W. Z. Jiang and B. A. Li, Mod. Phys. Lett. A \textbf{23}  (2008) 3393.

%\bibitem{ref29}K. Lim, C. J. Horowitz, Nucl. Phys. A \textbf{490}, 729 (1988).
\bibitem{ref27}D. Vretenar, A. Wandelt, P. Ring, Phys. Lett. B \textbf{487}  (2000) 334.



\bibitem{ref28}Z-Y. Ma, N. V. Giai,  A. Wandelt, D. Vretenar, P. Ring, Nucl. Phys. A \textbf{686}   (2001) 173.
\bibitem{ref29}P. Ring, Z-Y. Ma, N. Van Giai, D. Vretenar, A. Wandelt, L-G. Cao, Nucl. Phys. A \textbf{694}  (2001) 249.
\bibitem{We68}S. Weinberg, Phys. Rev. \textbf{166}  (1968) 1568.
\bibitem{We79}S. Weinberg, Phys. A \textbf{96}  (1979) 327.
\bibitem{Ko97}V. Koch,  	Int. J. Mod. Phys. E \textbf{6}  (1997) 203.


\bibitem{ref37}W. G. Paeng, T. T. S. Kuo, H. K. Lee, and M. Rho, Phys. Rev. C \textbf{93}  (2016) 055203.
\bibitem{ref38}J. R. Pel\'aez, Phys. Rep. \textbf{658}  (2016) 1.
\bibitem{ref39}J. R. Pel\'aez, A.Rodas, and Ruiz de Elvira, J. Eur. Phys. J. Spec. Top. \textbf{230}   (2021) 1539.

\bibitem{Wo22}R. L. Workman, V. D. Burkert,  V. Crede,  E. Klempt,
 U. Thoma, et al. (Particle Data Group), Prog. Theor. Exp. Phys. \textbf{2022}  (2022) 083C01.

\bibitem{ref35}J. A. Oller, E. Oset, and J. R. Pelaez, Phys. Rev. Lett. \textbf{80} (1998) 3452.
\bibitem{ref36}G. Colangelo, J. Gasser, and H. Leutwyler, Nucl. Phys. B \textbf{603} (2001) 125.

\bibitem{Bri17}R. A. Briceno, J. J. Dudek, R. G. Edwards, and D. J. Wilson, Phys. Rev. Lett. \textbf{118}  (2017) 022002.




\bibitem{Abb17}B. P. Abbott, R. Abbott, T. D. Abbott, M. R. Abernathy, K. Ackley, C. Adams, P. Addesso, R. X. Adhikari, V. B. Adya, C. Affeldt et al., Class. Quan. Grav. \textbf{34}  (2017) 044001.
%\bibitem{ref33}A. Faessler, T. Gutsche, M.A.Ivanov, V. E. Lyubovitskij and P.Wang, Phys. Rev. D \textbf{68} 014011 (2003).
%\bibitem{ref34}A. V. Friesen, Y. L. Kalinovsky and V. D. Toneev, Particles and Nuclei Letters. 9 1 (2012)

%\bibitem{ref36}J. R. Morones-Ibarra and A. Santos-Guevara Acta Phy. Pol. B38 2555 (2007)
\bibitem{An12}N. Andersson, K. Glampedakis, W. C. G. Ho, C. M. Espinoza, Phys. Rev. Lett. \textbf{109} (2012) 241103.
\bibitem{Ch13}N. Chamel, Phys. Rev. Lett. \textbf{110} (2013) 011101.
\bibitem{Bo77}J. Boguta and A. R. Bodmer,   Nucl. Phys. A \textbf{292} (1977) 413.
\bibitem{St14}J. R. Stone, N. J. Stone, and S. A. Moszkowski, Phys. Rev. C \textbf{89} (2014) 044316.
\bibitem{Ro18}X. Roca-Maza, N. Paar, Prog. Part.  Nucl. Phys. \textbf{101} (2018) 96.

\bibitem{He13}K. Hebeler, J. M. Lattimer, C. J. Pethick, and A. Schwenk, Astrophys. J. \textbf{773} (2013) 11.

\bibitem{ref42}H. Kurasawa, T. Suzuki, Nucl. Phys. A \textbf{490} (1988) 571.
\bibitem{Rj88}R. J. Furnstahl, C. J. Horowitz, Nucl. Phys. A \textbf{485} (1988) 632.

\bibitem{Li08}B. A. Li, L. W. Chen, and Che Ming Ko, Phys. Rep. \textbf{464} (2008) 113.
\bibitem{Jm16}J. M. Lattimer, and  M. Prakash, Phys. Rep. \textbf{621} (2016) 127.



\bibitem{ref43a}M. Oertel, M. Hempel, T. Kl\"ahn, and S. Typel, Rev. Mod. Phys. \textbf{89} (2017) 015007.

\bibitem{ref43}A. Carbone, G. Col\`o, A. Bracco, L-G. Cao, P. F. Bortignon, F. Camera, and O. Wieland, Phys. Rev. C \textbf{81}  (2010) 041301(R).
\bibitem{ref44}M. B. Tsang, J. R. Stone, F. Camera,  P. Danielewicz, S. Gandolfi, K. Hebeler, C. J. Horowitz et al., Phys. Rev. C \textbf{86} (2012) 015803.

\bibitem{ref45}J. M. Lattimer and Y. Lim, Astrophys. J. \textbf{771}  (2013) 51.           %63
\bibitem{ref47}B. A. Li and  X. Han,  Phys. Lett. B \textbf{727}  (2013) 276.            %64
\bibitem{ref48}J. Xu, W. J. Xie, and B. A. Li, Phys. Rev. C \textbf{102} (2020) 044316.  %65
\bibitem{ref49}W. G. Newton and G. Crocombe, Phys. Rev. C \textbf{103} (2021) 064323.   %66

\bibitem{ref50}B. T. Reed, F. J. Fattoyev, C. J. Horowitz, and J. Piekarewicz, Phys. Rev. Lett. \textbf{126} (2021) 172503. %27  67

\bibitem{PREX}D. Adhikari, H.  Albataineh, D. Androic, K. Anio, et al.  Phys. Rev. Lett. \textbf{126}   (2021) 172502.  %26         68


\bibitem{Roca15}X. Roca-Maza, X. Vi\~nas, M. Centelles, B. K.  Agrawal, G. Col$\grave{\rm o}$,  N. Paar, J. Piekarewicz, D. Vretenar,  Phys. Rev. C \textbf{92} (2015) 064304.                                %34
\bibitem{ref54}M. Oertel, M. Hempel, T. Kl$\ddot{a}$hn, and S. Typel, Rev. Mod. Phys. \textbf{89} (2017) 015007. %69   33
\bibitem{ref55}J. Estee, W.G. Lynch, C.Y. Tsang, J. Barney et al. Phys. Rev. Lett. \textbf{126} (2021) 162701. %70     28




\bibitem{Mo10}Ch. C. Moustakidis, T. Niksi\'c, G. A. Lalazissis, D. Vretenar, and P. Ring, Phys. Rev. C \textbf{81} (2010) 065803.

\bibitem{Sh21}W. Z. Shangguan, Z. Q. Huang, S. N. Wei,
and W. Z. Jiang,     Phys. Rev. D \textbf{104} (2021) 063035.


\bibitem{Ab18}B. P. Abbott et al. (LIGO Scientific Collaboration and Virgo Collaboration), Phys. Rev. Lett. \textbf{121} (2018)  161101.


\bibitem{We11}S. Weissenborn, D. Chatterjee, and J. Schaffner-Bielich, Nucl. Phys. A \textbf{881} (2012) 62.

\bibitem{De10} P. Demorest, T. Pennucci, S. Ransom, M. Roberts and J. Hessels, Nature \textbf{467}(2010) 1081.

\bibitem{An13} J. Antoniadis, P. C. Freire, N. Wex, T. MTauris, R. S. Lynch, et al., Science \textbf{340}  (2013) 1233232.
\bibitem{Hai08} M. M. Haidari and M. M. Sharma, Nucl. Phys. A \textbf{803} (2008) 159.
\bibitem{Sha09}G-Y. Shao and Y-X. Liu, Phys. Rev. C  \textbf{79} (2009) 025804.
\end{thebibliography}
\end{document}